\documentclass{nature}

\usepackage[utf8]{inputenc}
\usepackage{xcolor}
\usepackage{amsmath}
\newcommand{\Rev}[1]{\textcolor{black}{ #1}}
\newcommand{\RevF}[1]{\textcolor{black}{ #1}}

\usepackage{braket}
\usepackage{enumitem}
\usepackage[mathscr]{euscript}
\usepackage{graphicx}
\usepackage{caption}
\usepackage{enumitem}
\usepackage{lineno}
\usepackage{comment}
\usepackage{fixltx2e}

\makeatletter
\let\saved@includegraphics\includegraphics
\AtBeginDocument{\let\includegraphics\saved@includegraphics}
\renewenvironment*{figure}{\@float{figure}}{\end@float}
\makeatother

\title{Free-standing bilayer metasurfaces in the visible}

\author{Ahmed H. Dorrah$^{1,2,\dagger}$, Joon-Suh Park$^{1,\dagger}$, Alfonso Palmieri$^{1,\dagger}$ and Federico Capasso$^{1,*}$} 

\begin{document}
\captionsetup[figure]{labelfont={bf},name={Figure},labelsep=colon,font=small}

\maketitle

\begin{affiliations}
 \item John A. Paulson School of Engineering and Applied Sciences, Harvard University, Cambridge, Massachusetts 02138, USA \\
\item Department of Applied Physics and Science Education, Eindhoven University of Technology, Eindhoven 5612 AP, The Netherlands\\

 $^\dagger$These authors contributed equally\\

{*}Corresponding author(s): capasso@seas.harvard.edu
\end{affiliations}

\begin{abstract}
Mult-layered meta-optics have enabled complex wavefront shaping beyond their single layer counterpart owing to the additional design variables afforded by each plane. For instance, complex amplitude modulation, generalized polarization transformations, and wide field of view are key attributes that fundamentally require multi-plane wavefront matching. Nevertheless, existing embodiments of bilayer metasurfaces have relied on configurations which suffer from Fresnel reflections, low mode confinement, or undesired resonances which compromise the intended response. Here, we introduce bilayer metasurface\RevF{s} made of free-standing meta-atoms working in the visible spectrum. We demonstrate \RevF{their} use in wavefront shaping of linearly polarized light using pure geometric phase with diffraction efficiency of 80 \% --- expanding previous literature on Pancharatnam-Berry phase metasurfaces which \RevF{rely on} circularly or elliptically polarized illumination. The fabrication relies on a two-step lithography and selective develop\RevF{ment} processes which yield free standing, \RevF{bilayer} stacked metasurfaces, of 1200 nm total thickness. The metasurfaces comprise TiO\textsubscript{2} nanofins with vertical side walls. Our work advances the nanofabrication of compound meta-optics and inspires new directions in wavefront shaping, metasurface integration, and polarization control.      

\end{abstract}

\section{Introduction}

\begin{figure}[hb!]
    \centering
    \includegraphics[width=0.975\textwidth]{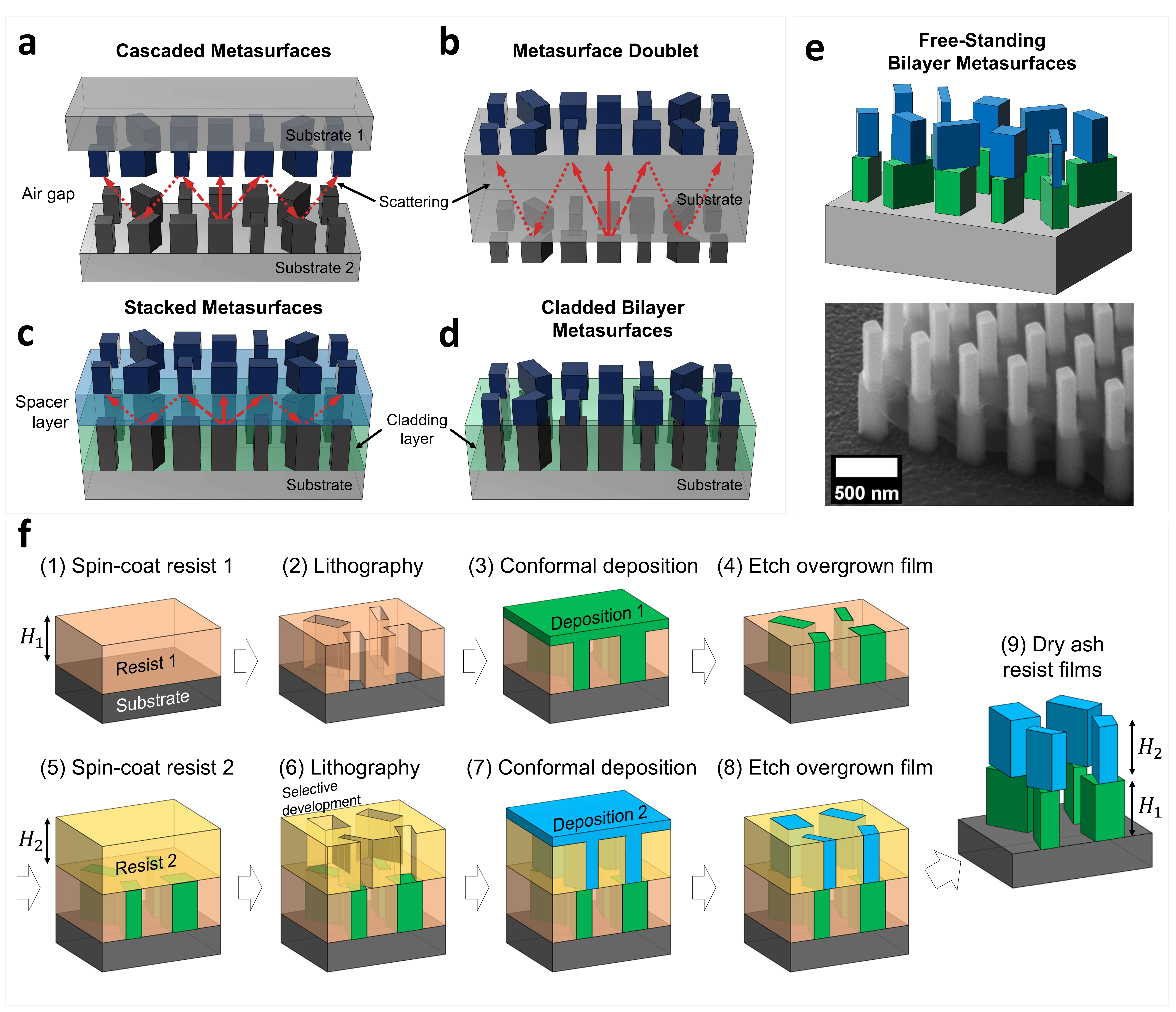}
    \caption{\textbf{Schematics of multi-layer metasurface configurations}. \textbf{a)} Cascaded metasurfaces are realized by stacking two or more metasurfaces either in contact or with an air gap in between. Shown (red arrows) is undesirable light scattering arising between the two layers. \textbf{b)} Double sided metasurfaces are created by patterning meta-atoms on the front and back side of the substrate. In cladded bilayer metasurfaces, the bottom meta-atoms are surrounded by a protective polymer to support the top layer which can be separated (\textbf{c)}) or in direct contact with the bottom layer \textbf{d)} . \textbf{e)} Free standing bilayer metasurfaces consist of two layers of meta-atoms surrounded by air thereby achieving high index contrast and mode confinement within the nanofins. An SEM image of a bilayer TiO\textsubscript{2} metasurface fabricated with our process is shown at the bottom. \textbf{f)} Fabrication process for free-standing bilayer metasurfaces.}
    \label{Fig1}
\end{figure}

Flat optics has emerged as a versatile wavefront shaping tool due to its sub-wavelength resolution, ease of integration, and compact footprint~\cite{doi:10.1126/science.1210713,doi:10.1126/science.1232009,doi:10.1126/science.1253213,Yu2014,Genevet:17}. Composed of sub-wavelength spaced arrays of nanoscatterers, metasurfaces have radically transformed the capabilities of conventional lenses~\cite{https://doi.org/10.1002/adom.201800554, Pan2022} and have enabled extreme control over light's degrees-of-freedom both in space and time~\cite{PhysRevLett.125.093903, KRASNOK20188, 10.1093/nsr/nwy017, doi:10.1126/science.aax2357, doi:10.1126/science.aav9632, Kuznetsov2024}. The multi-functionality~\cite{KamaliArbabiFaraon, doi:10.1126/science.abi6860} and tunability~\cite{doi:10.1126/science.aat3100, Yang_2022} of metasurfaces hold the promise for addressing many challenges in biomedical imaging~\cite{ZhangWongZengBiTaiDholakia,Pahlevaninezhad2022} and sensing~\cite{Yesilkoy2019,JulianoMartins2022}, fiber-based communications~\cite{Oh2022,oh2023metasurfaces}, space domain awareness~\cite{Cheng:14} and AR/VR applications~\cite{Lee2018,Song2021}, to name a few. Applications of this kind often deploy a single layer metasurface as a planar phase mask to alleviate the complexity associated with the fabrication of 3D metamaterials in the visible. Nevertheless, as metasurface-enabled technologies started to mature, the quest for more advanced functionality also evolved creating a growing demand for more complex configurations such as cascaded~\cite{doi:10.1126/sciadv.abf9718,Zhou2019,Huang:19}, double sided~\cite{Arbabi2016,Groever2017}, and multi-layered flat optics~\cite{https://doi.org/10.1002/adma.202204085} (Fig. \ref{Fig1}(a)).

Compound meta-optics have emerged in part because a single interaction between light and a flat optic is fundamentally limited to a finite set of allowable functions, \Rev{i.e., optics need thickness~\cite{doi:10.1126/science.ade3395}}. For instance, one cannot align two optical beams collinearly with a single mirror. Spatial mode multiplexing, which requires simultaneous wavefront shaping and translation (lateral shift) of an incoming beam array into a collinear beam, is similarly a task that requires multi-plane light conversion~\cite{Oh2022}. Likewise, arbitrary complex amplitude modulation with high efficiency cannot be realized with a single dielectric metasurface~\cite{PhysRevX.6.041008}. This can be understood from a simple input/output local power matching point-of-view and thus requires cascading at least two layers to achieve lossless polarization and phase conversion~\cite{https://doi.org/10.1002/adom.202102591,Zheng2022}. In general, distributing a complex wavefront transformation onto multiple layers provides an additional degree-of-freedom (i.e., desirable redundancy) in metasurface design which can be used to achieve multi-functionality, broadband operation, wide field of view, and versatile dispersion control~\cite{Chen2020}. From \RevF{a} polarization optics standpoint, shape-birefringent dielectric nanofins mimic the function of wave plates whose transmission function acquires the form of a 2-by-2 unitary and symmetric Jones matrix~\cite{Rubin:21}. While such polarization control has enabled single shot Stokes and Mueller imaging~\cite{doi:10.1126/science.aax1839, Zaidi2024}, exotic classes of polarizers~\cite{Dorrah2021}, and vectorial holograms~\cite{doi:10.1126/sciadv.abg7488,Yue2016}, it fails to realize more general functionalities. For example, a single layer non-chiral nanofin cannot realize a circular analyzer by relying on its shape-birefringence alone as this requires the off-diagonal terms of the Jones matrix to be decoupled. In this pursuit, bilayer metasurfaces have been deployed to enable more general polarization transformations~\cite{https://doi.org/10.1002/adma.202204085,Bao2022} by decoupling all four elements of the Jones matrix. This hinges on the mathematical fact that a product of two symmetric matrices can yield an arbitrary matrix that is not necessarily symmetric~\cite{PhysRevLett.129.167403,Palmieri:24}.

Despite the wide interest in compound or cascaded meta-optics, their previous implementations have been limited to stacking multiple metasurfaces (Fig.~\ref{Fig1}(a)), patterning meta-atoms on both sides of a substrate (often referred to as metasurface doublets, Fig.~\ref{Fig1}(b)), or creating bilayer structures embedded in a cladding as depicted in Fig.~\ref{Fig1}(c-d). The approaches shown in Fig.~\ref{Fig1}(a-c) suffer from undesired Fresnel reflections and Fabry-Perot resonances whereas the approaches in Fig.~\ref{Fig1}(c-d) exhibit lower index contrast between the nanostructures and the surrounding environment. The latter subsequently reduces the allowable phase coverage and causes undesired coupling between adjacent meta-atoms, compromising the intended response. To evade the optical losses and unwanted reflections caused by the embedding material or the gap between the cascaded metasurfaces, it is evident that free-standing, stacked meta-atoms in direct contact with each other would be desirable. 

Here, we demonstrate the bilayer metasurfaces composed of free-standing titanium dioxide (TiO\textsubscript{2}) nanofins directly stacked on top of one another, operating in the visible spectrum (Fig. \ref{Fig1}(b)). Each nanofin is $600$ nm in height, enabling independent $0-2\pi$ phase coverage at each layer, achievable by having a high index contrast with respect to its surroundings. As an example of demonstrating the versatility and applicability of such bilayer metas-optics, we realize a reflecting bilayer meta-optics that can impart a geometric (\textit{i.e.}, Pancharatnam-Berry) phase~\cite{Cohen2019}, point-by-point, on a linearly polarized light basis. This is in contrast to the wide class of geometric phase metasurfaces in previous literatures, which are dependent on circularly or elliptically polarized light incidences\cite{yuan_reaching_2024,shi_freeform_SciAdv2020}. Our approach inspires new directions in nanophotonics fabrication and brings the broadband operation, high-efficiency, and robustness of geometric phase flat optics to linear polarization basis which can advance many applications in classical and quantum imaging, sensing, and optical communications.   

\section{Concept and Methodology}

\subsection{Fabrication Method}

As illustrated and replicated in various studies, single-layer TiO\textsubscript{2} meta-atoms can be reliably fabricated using a damascene-like process.\cite{devlin_pnas,Chen2020,Rubin:21,shi_freeform_SciAdv2020,chen_dispersion-engineered_2023} In detail, an electron-beam (e-beam) resist film with a thickness $H_1$ is patterned to have vertical holes in the shapes of meta-atoms. For positive-tone resists, the cross-linked chemical bonds in the resist film are broken by bombarding high kinetic energy e-beams through the film in desired shapes, which transforms the exposed film into areas with broken bonds soluble by a developing solvent. By tuning the e-beam energy and the developing conditions, one can achieve various sidewall tapering profiles in the resist. Here, we focus on vertical sidewall structures for simplicity. The areas with broken chemical bonds by high-energy e-beam are then washed away in a developer solvent, creating meta-atom-shaped holes in the resist film (Resist 1 in Fig.\ref{Fig1}(f), step (2), Figs. S1-S2).

However, during the e-beam write on the top layer, the resist on the bottom layer also gets exposed to ballistic \RevF{electrons}; both exposed area on top and bottom resists become soluble to a developer solution. Therefore, the choice of the resist and the developer solution becomes critical: Although both the resist layers are simultaneously exposed, if the exposed bottom layer resist chemicals are insoluble or have very low solubility to the developer solution for the top layer resist, only the exposed top layer becomes developed while the bottom layer resist film remains intact. Here, we choose ZEP520A (\textit{Zeon SMI}) and o-Xylene (puriss. p.a., $\geq99\%$ (GC), \textit{Sigma Aldrich}) as a \RevF{top} resist layer and its developer, respectively, and PMMA (950 PMMA A7,\textit{Kayaku Advanced Materials Inc.}) and MIBK/IPA 1:3 Positive Radiation Developer (\textit{Kayaku Advanced Materials Inc.}) or H\textsubscript{2}O/IPA 1:3 solution for the \RevF{bottom} resist layer and its developer, respectively (Fig. S1-S3).\cite{rooks_low_2002} This choice of resist and developer sets allows selective development of the top layer resist while minimally affecting the bottom layer resist (Fig.\ref{Fig1}(f), step (6), Fig. S4).  

After the top layer meta-atom shapes are defined in the top layer of resist as holes, a second ALD process is performed to fill the patterns with \RevF{TiO\textsubscript{2}} (Fig.\ref{Fig1}(f) step (7)). Then, the over-grown dielectric film is then etched away, until the top layer resist is exposed. The resulting feature is shown in Fig.~\ref{Fig1}(f) step (8), where the two dielectric nanostructures are stacked on top of each other and surrounded by resist layers. The resists are then selectively removed using remote oxygen plasma ashing, a dry etching process that allows solvent-free, gentle removal of the resist films while preserving the high-aspect-ratio structures, which are prone to damage from surface tension when exposed to solvents. (Fig.\ref{Fig1}(f) step (9)). An exemplary scanning electron microscope (SEM) image of fabricated bilayer nanostructures is shown in the bottom of Fig.\ref{Fig1}(b). More details on the fabrication process can be found in the Supplementary Information. We note that the two stacked nanostructures can be comprised of various materials that are compatible with the conformal coating process such as hafnia, zirconia, silica, alumina, zinc oxide, platinum, copper, etc. In addition, the structures do not necessarily need to be simple pillar or rectangular shapes but also can be complex shapes\cite{chen_dispersion-engineered_2023,shi_freeform_SciAdv2020} that could open grounds for more complex polarization or optical dispersion control for metasurface optics. In the following, we demonstrate two geometric phase metasurfaces which are realized using this fabrication process.

\subsection{Bilayer Geometric Phase Metasurface} 

 \begin{figure}[!hb]
    \centering
    \includegraphics[width=0.995\textwidth]{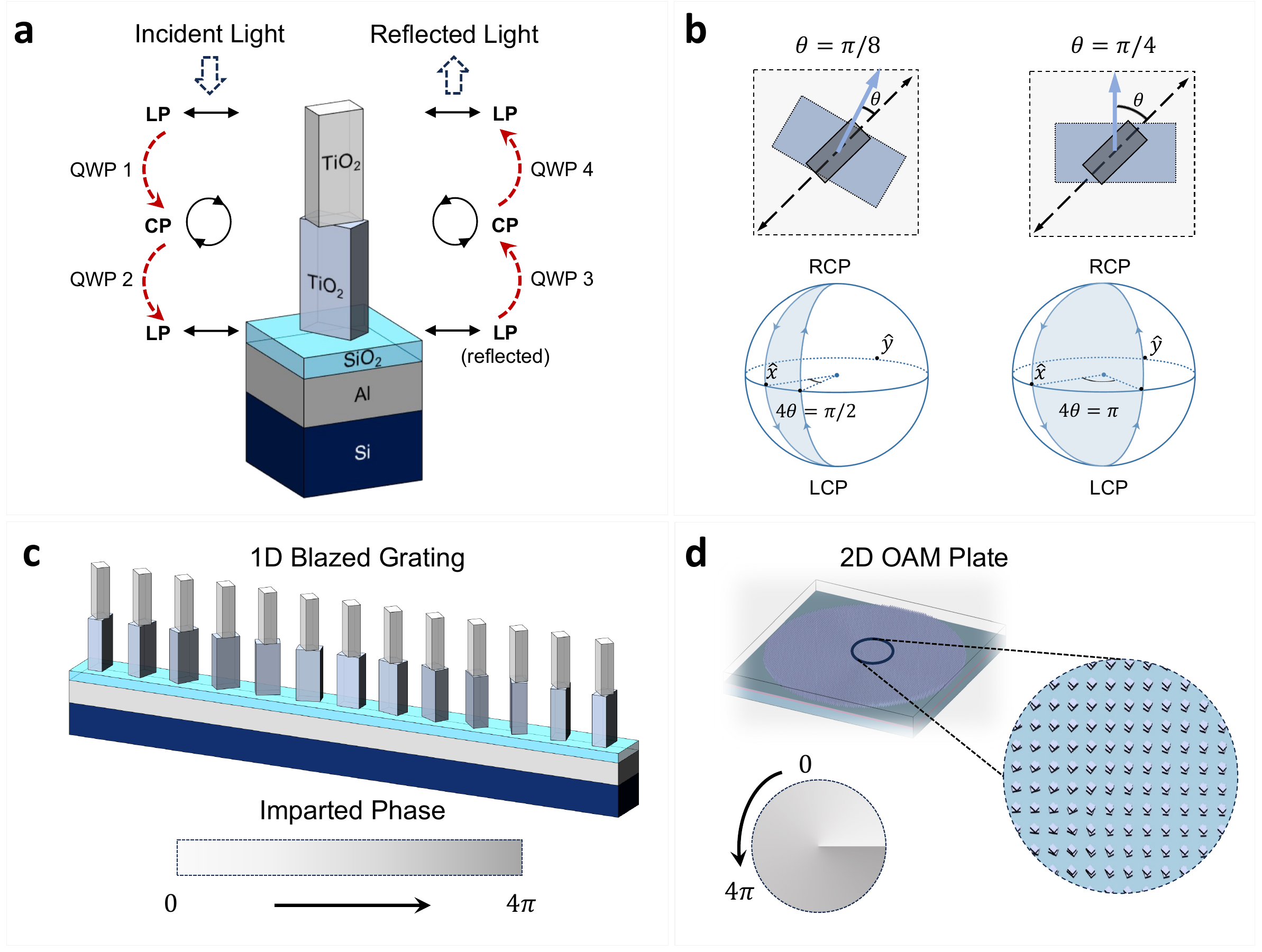}
    \caption{\textbf{Concept of bilayer PB phase metasurface operation}. \textbf{a)}. Two quarter-wave plate-like nanofins are stacked on top of one another and placed on an aluminum mirror with a thin SiO\textsubscript{2} spacer in between. The top and bottom nanofins are rotated at an angle of $\pi/4$ and $\pi/4 + \theta$ with respect to the polarization of incoming light. When x-polarized light interacts with the top nanofin, it becomes left handed circularly polarized, and then impinges on the bottom nanofin which in turn converts this circular polarization back to linear. The output light bounces off the aluminum mirror and passes through the two nanofins in reverse, changing its polarization from linear to right hand circular then back to linear at the output of the top nanofin. \textbf{b)} The path traversed by x-polarized light visualized on the Poincaré sphere as a function of the relative rotation angle ($\theta$) between the top and bottom nanofins. A rotation angle $\theta$ gives rise to a solid angle of $4\theta$, thereby imparting a geometric phase of $\pm2\theta$ on x- and y-polarized light, respectively. By locally changing the relative rotation angle between the top and bottom nanofins, point-by-point, across the metasurface, a 1D blazed grating (\textbf{c}) and a 2D vortex plate (\textbf{d}) can be designed.}
    \label{Fig2}
\end{figure}

Pancharatnam-Berry (PB) - or geometric - phase metasurfaces represent a wide class of birefringent flat optics which can impart $0-2\pi$ phase shift, pixel-by-pixel, on right/left circularly polarized light by varying the angular orientation of half-wave plate meta-atoms~\cite{Cohen2019,https://doi.org/10.1002/lpor.202100003,Rubin:21,doi:10.1073/pnas.2122085119, Khorasaninejad2014, PhysRevLett.96.163905, Bomzon:02}. More generally, when light passes through a sequence of polarizing elements, causing its original state of polarization to traverse a cyclic trajectory on the Poincaré sphere, the output beam acquires an additional phase shift governed by the topology of the path traversed in polarization space independent from the distance traversed in space. The curvature of the Poincaré sphere, which visualizes all possible states of polarization, allows this phase factor to be geometrically evaluated as half the solid angle enclosed by the traversed topological path. Generalizations of this rule that apply to nonadiabatic and/or noncyclic topological evolution have also been reported~\cite{PhysRevLett.60.2339,doi:10.1126/sciadv.aay8345}. To this end, previous geometric phase meta-optics have primarily operated on circularly polarized light, reversing its handedness at the output while locally imparting a geometric phase that is twice the angular orientation of each nanofin. Here, we show \RevF{instead} that stacking two nanofins on top of one another---thanks to the fabrication process proposed above---enables a class of geometric phase flat optics which performs wavefront shaping on linearly polarized light instead.

Figure \ref{Fig2}(a) depicts the building block of our metasurface; a stack of two quarter-wave plate (QWP) nanofins made of titanium dioxide (TiO\textsubscript{2}), standing atop of a 150 nm thick Al reflection layer with a ~$90$ nm thick SiO\textsubscript{2} spacer layer in between. Consider a scenario in which linearly polarized (LP) light in x-axis direction impinges on the top nanofin. If the nanofin's principle axis makes an angle $\pi/4$ with the x-axis, then the transmitted light becomes left-hand circularly polarized (CP) upon exiting from the top nanofin. In this case, the polarization state has traversed a continuous path on the Poincaré sphere from its equator towards the south pole. When this circularly polarized light passes through the bottom nanofin (whose slow axis makes an angle of $3\pi/4 + \theta$ with the x-axis), it gets converted back to linear polarization, located on the equator of the Poincaré sphere. This beam reflects off a plane mirror and passes back through the two quarter-wave plates, converting its polarization from linear to right hand circular then back to linear in an adiabatic manner. Hence by interacting with this bilayer meta-atom in reflection, the emerging beam has traversed a closed circuit on the Poincaré sphere as illustrated in Fig.~\ref{Fig2}(b). The solid angle of the traversed path is equal to $4\theta$ which implies that a PB phase of $2\theta$ is acquired by x-polarized light. For y-polarized incident illumination, the enclosed path is still cyclic but traverses in the opposite direction, giving rise to an accumulated phase factor of $-2\theta$. 

To realize the bilayer meta-atoms, we had studied the evanescent coupling between the top and bottom nanofins and explored the domains over which this coupling can be neglected. Our analysis suggest that the coupling between the nanofins is negligible if neither of the two nanofins are in resonance and if the top nanofin has smaller dimensions than the bottom one~\cite{Palmieri:24}. Accordingly, we selected two quarter-wave plate-like nanofins which satisfy these criteria and utilized them as the building blocks of our bilayer metasurface. Additionally, since the aluminum layer is not a perfect mirror (\textit{i.e.}, it introduces a complex phase shift to impinging light), we added a thin matching layer of a dielectric spacer made of ~$90$ nm thick SiO\textsubscript{2} below the meta-atom. The bottom and top nanofins are $134$-by-$202$ nm and $114$-by-$154$ nm, respectively, with a \RevF{unit cell size} of $420$ nm \RevF{(Fig. \ref{Fig2}(a))}. This choice of dimensions maximizes the overlapping surface area connecting the two nanofins, ensuring their structural stability. Using this meta-atom configuration, we designed a 1D reflective blazed grating that deflects linearly polarized light in x- or y-axis to -1 or +1 order, respectively, and a 2D reflective orbital angular momentum (OAM) plate which converts a Gaussian beam into a vortex beam with a helical wavefront of $0-4\pi$ along the azimuthal direction, giving rise to a topological charge of $\ell=\pm 2$. Full scale simulations of both devices have been performed using a commercial finite-difference time-domain simulation tool (Tidy 3D). In the following, we present SEM images and optical characterization of the fabricated devices.

\section{Results}

\subsection{Experimental characterization}

 \begin{figure}[!ht]
    \centering
    \includegraphics[width=0.975\textwidth]{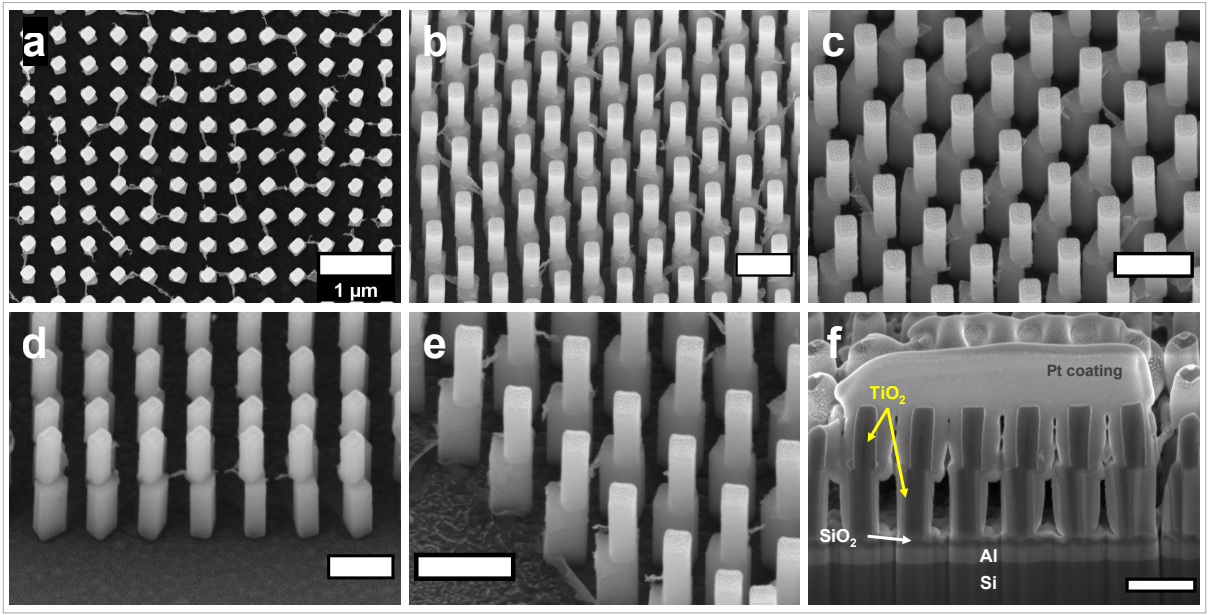}
    \caption{\textbf{Fabricated free-standing bilayer metasurface}. \textbf{a)} Scanning electron microscope (SEM) image of the fabricated bilayer metasurface viewed from the top.  (b-e) Tilted-view SEM images of the fabricated bilayer metasurfaces. Scale bars: 500 nm. (f) Cross-sectional SEM image of the fabricated sample obtained by focused-ion beam (FIB) milling. The SiO\textsubscript{2} spacer layer and the Al mirror layer are visible in the image. Thick Pt coating was performed on the region of interest to protect the underlying features from FIB milling process. Scale bar: 500 nm.} 
    \label{Fig3}
\end{figure}


We designed and fabricated a metasurface which imparts a blazed grating profile --- i.e., a linear phase ramp of $0$-$2\pi$ --- with a grating periodicity of $5.04$ $\mu \text{m}$, spanning $500 \ \mu \text{m} \times 500 \ \mu \text{m}$ area. Figure \ref{Fig2}(c) depicts a periodic cell of the meta-grating comprising 12 meta-atoms, each composed of a bilayer nanofin. Within each meta-atom, the rotation angle between the top and bottom nanofins is varied from $0^\circ$ to $180^\circ$ with equal increments of $15^\circ$. As discussed in the previous section, a relative rotation angle of $\theta$ between the two nanofins imparts a geometric phase of $2\theta$. Hence, our 12-pixel unit cell enables a full phase ramp from 0 to $2\pi$. From the grating equation, the deflection angle for the $\pm1$ diffraction orders is $6.379^\circ$ at an incident wavelength of $560$ nm, where both the bottom nanofin and the top nanofin function as quarter-wave plates. 

Figure \ref{Fig3} shows SEM images of the fabricated free-standing bilayer metasurfaces operating in reflection to demonstrate the complex polarization control with bilayer meta-atoms. ZEP520A and PMMA resist was used to fabricate the bottom and top layer, respectively. Figure \ref{Fig3} (a) shows the top-view image of the blazed grating profile bilayer metasurface. The thin residual structures between the pillars are under-etched TiO\textsubscript{2} overgrown film during step (4) in Figure \ref{Fig1} (f). These fabrication imperfections minimally affect the optical performance of the device due to their small feature sizes and thickness ($<$ 10 nm). Figure \ref{Fig3} (b)-(e) show SEM images of the bilayer structures obtained at a $30^{\circ}$ tilted view. As designed (Fig. \ref{Fig2}(c)), it can be seen that the top layer nanofins are oriented in the same direction while the bottom layer nanofins are rotated with respect to each other. The scale bars represent 500 nm. Figure \ref{Fig3} (f) shows a cross-sectional SEM image of the fabricated bilayer metasurface obtained by focused-ion beam (FIB) milling. A thin Pt coating layer was applied to the region of interest to preserve and protect the structures against the high-energy ion beams during the milling process. The underlying reflective mirror comprising a 90 nm thick SiO\textsubscript{2} spacer layer and a 150 nm thick Al film on a Si substrate is visible at the bottom of the image. The scale bar represents 500 nm. 

\begin{figure}[!h]
    \centering
    \includegraphics[width=0.83\textwidth]{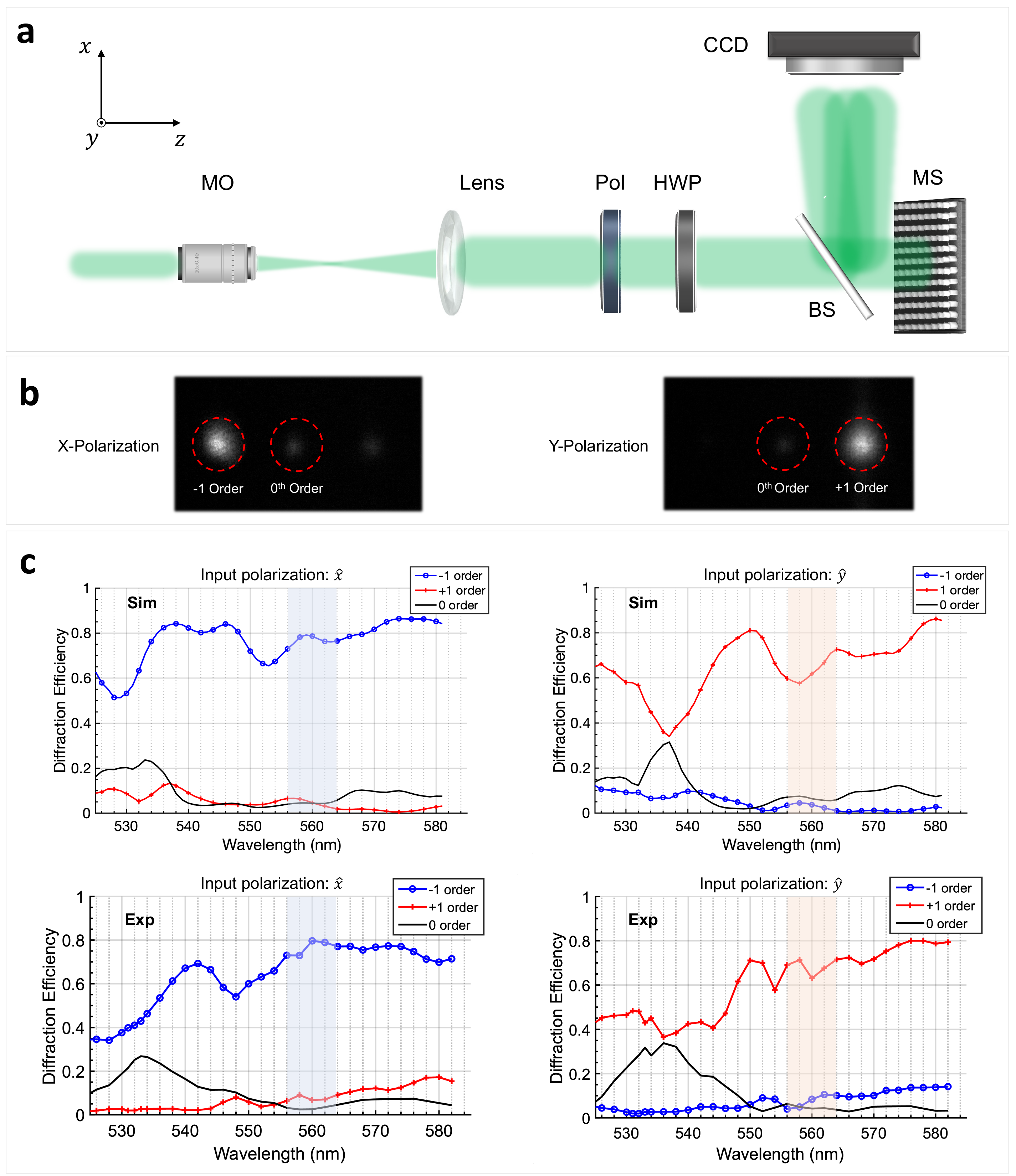}
    \caption{\textbf{Meta-grating characterization}.  \textbf{a)} Schematic of the experimental setup used to measure the metasurface response: a collimated beam from a supercontinuum source is expanded and collimated using a mircroscope objective (MO) and a lens. The polarization of the output beam is controlled using a half-wave plate (HWP) and a polarizer (Pol) before illuminating the metasurface (MS). A 50-50 beam splitter (BS) is used to redirect the light reflected from the metasurface onto a CCD or a power detector. \textbf{b)} Recorded images of the grating diffraction orders under x-polarization (left) and y-polarization (right). \textbf{c)} Simulated (top) and measured (bottom) diffraction efficiency as a function of wavelength under x- and y-polarized illumination. The diffraction efficiencies for the $-1$, $0$, and $+1$ orders are plotted for each case.}
    \label{Fig4}
\end{figure}

Figure \ref{Fig4}(a) depicts the experimental setup used to characterize the sample and measure the grating efficiency over a broadband. A wavelength-selective supercontinuum laser source (SuperK Extreme, \textit{NKT Photonics} with LLTF Contrast$^{TM}$, \textit{Photon etc.}) is expanded and collimated using a pair of a microscope objective (MO) and a refractive lens (Lens) before illuminating the metasurface (MS). A polarizer (Pol) and a half-wave plate (HWP) are used to control the polarization of the incoming beam; then a $50$-$50$ beam splitter (BS) is used to redirect the reflected diffraction order onto either a power detector or a CCD camera. 

The output response of this meta-grating is shown in Fig. \ref{Fig4}(b) where x-polarized light is deflected into the $-1$ order and y-polarized light into the $+1$ order, as expected. Repeating this measurement over a frequency range from $520$ nm to $580$ nm reveals the fairly broadband operation of our device, as shown in Fig. \ref{Fig4}(c). Here, the diffraction efficiency is defined as the total power deflected into the first order divided by the total power incident on the metasurface. We recorded diffraction efficiency of $80\%$ at $560$ nm under x-polarized illumination and a mean diffraction efficiency close to $60\%$ over the considered frequency range. This fairly broadband operation is a characteristic of geometric phase metasurfaces which are known for their robustness to fabrication tolerances. It is also achieved since the imparted phase stems from the evolution of the polarization state rather than the optical path length or material properties; hence, it is has a topological nature. Such response is preserved as long as the nanofins continue to function as quarter-wave plates, thereby maintaining the path traced on the Poincaré sphere shown Fig. \ref{Fig2}(b). 
\RevF{Because of the refractive index dispersion, the response of the nanofins deviates from that of a quarter wave plate away from the design wavelength $\lambda = 560$ nm (Fig. S5). This leads to a stronger zero-th order and to smaller diffraction efficiency, as observed at shorter wavelength in Fig.~\ref{Fig4}(f).}

\begin{figure}[h!]
    \centering
    \includegraphics[width=0.95\textwidth]{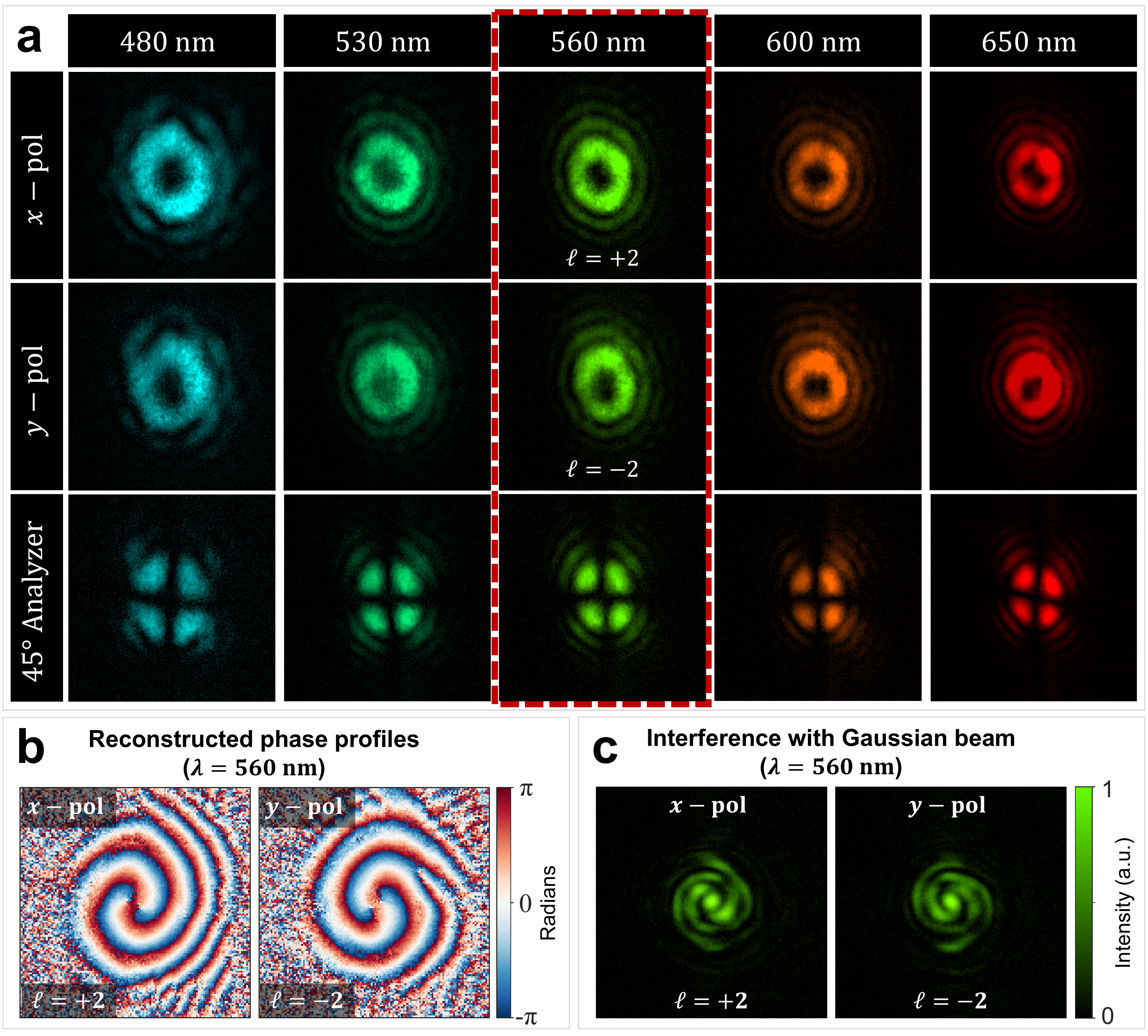}
    \caption{\textbf{\RevF{Vortex beam generation by the bilayer metasuface: experimental results}}. \textbf{a)} Intensity profiles of the orbital angular momentum (OAM) beams generated by the bilayer metasurface for different wavelengths. The first row shows the OAM mode with a topological charge of $\ell=+2$ when the incident polarization is along the x-axis. The second row represents the OAM mode with a topological charge of $\ell=-2$ for y-axis polarization. The third row shows the resulting petal-like interference patterns when a 45° analyzer is introduced \RevF{between} the beam splitter \RevF{and the camera}, causing the x and y components to interfere, demonstrating the characteristic petal structure indicative of the interference of OAM modes with opposite topological charges. The red dotted box marks the output response at the designed wavelength. \textbf{b)} Reconstructed phase profiles of the generated OAM for the x and y polarization. \textbf{c)} Measured spiral patterns created by the interference of the vortex beam and a co-propagating Gaussian beam. Depending on the sign of the topological charge $\ell=\pm2$, the helicity of the spiral is clockwise or counter-clockwise, respectively.}
    \label{Fig5}
\end{figure}

\RevF{W}e also created a 2D vortex plate which introduces an azimuthal phase ramp of $\exp(i\ell\phi)$ on incoming light, producing a vortex beam with orbital angular momentum (OAM)~\cite{Yao:11}. The latter stems from a non-zero transverse component of the Poynting vector owing to the helical phasefront which gives rise to a longitudinal component of OAM of $\ell\hbar$ per photon~\cite{ALLEN1999291}. The OAM plate is designed by varying the relative rotation angle between the two nanofins in each unit cell from $0$ to $2\pi$ along the azimuthal direction across the sample (Fig. \ref{Fig2}(d)). In this case, an azimuthal phase dependence of $0$-$4\pi$ is imparted on the incoming beam. The output intensity and phase response of this device is shown in Fig.~\ref{Fig5} under x- and y-polarized light. Unlike single-layer PB phase OAM plates~\cite{PhysRevLett.96.163905}, which generate a conjugate pair of vortex beams in response to circularly polarized incidence states via spin-orbit coupling, our device imparts a pure geometric phase on linearly polarized light. Specifically, it converts x-polarized light into a vortex beam with $\ell=+2$ and y-polarized light into a vortex beam with $\ell=-2$. Figure~\ref{Fig5}(a) confirms the wavelength-agnostic response of our device over a range of $200$ nm in the visible spectrum. The Figure showcases two-dimensional intensity distributions captured for orthogonal linear polarizations across different wavelengths. The third row reveals the interference pattern that emerges if an analyzer, rotated by $45^{\circ}$, is introduced \RevF{between the beam splitter and the sensor} when illuminated by $45^{\circ}$ polarized light. This enables the interference between two OAM beams with topological charges of $\ell = \pm2$, giving rise to the characteristic petal-like pattern.

The retrieved phase of the output vortices under $560$ nm linearly polarized light is shown in Fig.~\ref{Fig5}(b), \Rev{exhibiting an intertwined pair of helical wavefronts with opposite handedness, corresponding to $\ell=\pm2$, in response to x- and y-polarized light}. The phase profiles were obtained using single-beam multiple-intensity reconstruction, which was applied to the 2D intensity map measured at various distances from the sample~\cite{pedrini2005wave}. \Rev{This is an iterative multi-plane phase retrieval algorithm which estimates the 2D phase profile of a given light intensity distribution propagating between two or more planes, in analogy to the Gerchberg Saxton algorithm. Technical details about the axial separation between the planes and the propagator are discussed more fully in the Supplemental Information.}  

\Rev{To further confirm the topological charge of the output beams, we carried out another interferomteric measurement. Figure~\ref{Fig5}(c) shows the characteristic spiral fringes obtained by interfering the vortex beam with a collinear Gaussian illumination. We realize this by collimating the illumination beam and expanding its aperture \RevF{to a} size comparable \RevF{to that of the} OAM plate. This creates strong background Gaussian illumination in the far-field which interferes with the output beam, producing the shown spiral patterns of opposite handedness (marking the sign reversal) and with two arms (indicating the topological charge magnitude), altogether confirming the generation of an optical vortex pair with topological charge of $\ell=\pm2$ in good agreement with Fig.~\ref{Fig5}(b).}

\section{Discussion and Outlook}
We introduced a new class of flat optics which consists of two free-standing nanofins stacked on top of one another and surrounded by air. The high index contrast afforded by each layer enables full and independent $0$-$2\pi$ phase coverage--- a redundancy that can be used for broadband operation and multi-functional design. Our fabrication protocol relies on selective development by judiciously choosing two sets of e-beam resist and developer for patterning the bottom and top layers, sequentially, with minimum interference with the previously fabricated layer. Although we demonstrated bilayer metasurfaces, our approach can be generalized to realize multi-layer flat optics by alternating the order of the two compatible sets of resists and developers. Using this fabrication method, we demonstrated TiO\textsubscript{2} bilayer geometric phase metasurfaces which operate on linear polarization basis. The underlying mechanism relies on allowing light's polarization to traverse a cyclic path on the Poincaré sphere whose solid angle depends on the relative rotation between the top and bottom nanofins. 

Based on this concept, we presented two widely used class of wavefront shaping devices which impart a phase gradient in 1D and 2D, respectively. The first is a blazed grating which deflects x- and y-polarized light by $\pm 15^\circ$ and the second is an OAM plate which creates vortex beams with a topological charge of $\pm2$ in response to x- and y-polarized light. In these devices, we recorded diffraction efficiencies of up to $80\%$ with fairly broadband operation which are well known characteristics of geometric phase metasurfaces. Although we adopted TiO\textsubscript{2}, other material platforms and free-form meta-atoms are also compatible with our approach which brings these capabilities to the IR and telecom wavelengths, or even for manufacturing complex vertical interconnects between modern multi-layered semiconductor circuits and photonic chips. We envision that this work will inspire new flat optics architectures which can advance polarization optics and wavefront shaping applications including holography, structured light, remote sensing, beam steering, and asymmetric transmission operations. 

\section*{Methods}
The mirror substrate was fabricated by first depositing a 150 nm thick Al film on a polished silicon substrate using e-beam evaporation (Sharon e-beam evaporator) followed by a deposition of 90 nm thick SiO\textsubscript{2} using a low-temperature inductively coupled plasma chemical vapor deposition (Oxford Instruments PlasmaPro 100 ICPCVD). E-beam lithography processes were performed using either 125 kV or 150 kV acceleration voltage EBL systems (Elionix ELS-F125, Elionix Boden-150). Atomic layer depositions of TiO\textsubscript{2} were performed using the reaction between Tetrakis(dimethylamido)titanium (TDMAT) precursor and water vapor at 90$^\circ C$ (Cambridge Nanotech Savannah). Remote plasma ashings were performed using Matrix 105 Plasma Asher. SEM images were obtained using Zeiss field emission SEM (FESEM) Ultra Plus, and FIB cross-section images were obtained using FEI Helios 660 FIB-SEM (Thermo Fisher Scientific).

\subsection{Acknowledgements}

We thank Jaewon Oh, Aun Zaidi, Noah A. Rubin, and Kerolos M. A. Yousef from Harvard University, as well as Paulo Dainese from Corning Inc. for their insightful discussions. We also acknowledge the help of staff from Harvard's Center for Nanoscale Systems (CNS); Stephan Kraemer with the FIB process, and Mac Hathaway with the ALD process. A.H.D. acknowledges financial support from the Optica Foundation Challenge program. F.C. acknowledges financial support from the Office of Naval Research (ONR) under the MURI program, grant no. N00014-20-1-2450, and from the Air Force Office of Scientific Research (AFOSR) under grant nos FA9550-21-1-0312 and FA9550-22-1-0243. This work was performed at the Harvard University Center for Nanoscale Systems (CNS); a member of the National Nanotechnology Coordinated Infrastructure Network (NNCI), which is supported by the National Science Foundation under NSF award no. ECCS-2025158. The simulation work was in part performed using Tidy3D software from Flexcompute.

\subsection{Author Contributions}
A.H.D. developed the theory of bilayer geometric phase metasurfaces, built the measurement setup, characterized and analyzed the data. J.-S.P. developed the fabrication processes for the bilayer metasurfaces, fabricated the devices, and acquired the SEM images. A.P. performed the simulation of the devices and contributed to their fabrication and characterization. F. C. supervised the project. All authors contributed to writing the manuscript.

\subsection{Data Availability}
All key data that support the findings of this study are included in the main article and its supplementary information. Additional data sets and raw measurements are available from the corresponding author upon reasonable request.

\subsection{Code Availability}
The codes and simulation files that support the figures and data analysis in this paper are available from the corresponding author upon reasonable request.


\subsection{Competing Interests} The author declare no competing interests.


\bibliography{Bibliography}


\end{document}